\documentclass[12pt,preprint]{aastex}

\def\cut{{\rm cut}}
\def\au{{\rm AU}}
\def\yr{{\rm yr}}
\def\kms{{\rm km\,s^{-1}}}
\begin{document}

\title{Searching for Failed Supernovae With Astrometric Binaries}

\author{Andrew Gould and Samir Salim} 
\affil{Department of Astronomy, The Ohio State University,
140 W.\ 18th Ave., Columbus, OH 43210}
\email
{gould,samir@astronomy.ohio-state.edu}

\singlespace

\begin{abstract}

Stars in the mass range $8\,M_\odot \la M \la 30\,M_\odot$ are thought to end
their lives as luminous supernovae that leave behind a neutron star.  However,
if a substantial fraction of these stars instead ended as {\it black-hole}
remnants, without producing a supernova (a `failed' supernova), how would one
know?  We show that, under plausible assumptions, the {\it Hipparcos} catalog
should contain $\sim 30\,f_{\rm fail}$ astrometric binaries with black-hole
companions, where $f_{\rm fail}$ is the fraction of supernovae that fail.
Since no black-hole astrometric binaries are found in {\it Hipparcos}, 
one might like to
conclude that such failed supernovae are very rare.  However, the most
important assumption required for this argument, the initial companion mass
function (ICMF) of G stars (the majority of {\it Hipparcos} stars) in the
high-mass companion regime, is without any observational basis.  We show how
the ICMF of G stars can be measured using the {\it Full-Sky Astrometric
Explorer (FAME)}, thereby permitting an accurate measurement of the rate of
supernovae that fail.

\end{abstract}
\keywords{astrometry---binaries: general---stars: neutron---supernovae:
general}
 
\section{Introduction
\label{sec:intro}}

        Massive stars ($M\ga 8\,M_\odot$) end their lives as supernovae (SNe),
leaving behind black-hole (BH) or neutron-star (NS) remnants depending on
whether they are more or less massive than a cutoff mass, $M_\cut\sim
30\,M_\odot$.  So says the standard lore, but what is the observational
evidence?

        Certainly there exist NSs and BHs, and since pulsars are frequently
found near the centers of SN remnants, there can be little doubt about their
origin.  But do the majority of massive stars $8\la M/M_\odot\la 30$ really
die gloriously in luminous SNe that give birth to a NS?  Or do most of them
fizzle out in failed SNe that collapse in on themselves, leaving behind a BH?

Failed SNe have been invoked by theorists for a number of reasons,
mostly because of the difficulties of producing an explosion in
analytical models and simulations (for a detailed review see
 \citealt{woos_araa}). The failures in the earliest models in the 1960s
 were mostly ascribed to the models' inability to transfer
 the energy of the core collapse into decoupling the mantle from the
 envelope. Improvements in the neutrino transport model and new
 estimates of neutrino cross sections still failed to produce
 explosions in the 1970s. Owing to improved nuclear rates and more
 sophisticated models, successful explosions were finally produced in
 the 1980s, however only for low-mass iron cores ($M\lesssim1.3\,M_\sun$),
 corresponding to progenitor masses of $M\lesssim 11\,M_\sun$. This
 problem was somewhat alleviated by the introduction of a delayed
 neutrino energy transport, which supplies the energy to the lower parts
 of envelope and so helps the shock propogate outwards. The success of
 this mechanism remains inconclusive. One should keep in mind that the
 researchers were driven to make explosions happen and not to
 demonstrate that they do not happen. \citet{mikaelian} early on
 suggested that some SNe really `fail', arguing that only stars that
 spin slowly can produce a SN, while the fast spinning
 stars just collapse into a BH. Others, however, have dismissed
 rotation as a significant factor.  Failed SNe (of type Ib)
 were suggested by \citet{woosley} to represent the main mechanism for
 producing cosmological $\gamma$-ray bursts (GRBs). In this conjecture,
 for which the simulations have already produced substantial
 support, 
the collapsing core fails to produce an explosion, but
 instead an accretion disk forms around the core, which draws in mass
 from the mantle at the rate of $\sim 0.5\,M_\sun\,{\rm s}^{-1}$, and then
quickly collapses into a BH.  \citet{woosley} suggests that in
 stars that have previously lost their H-envelope (Wolf-Rayet stars)
 the accretion could be accompanied by an energetic burst of gamma rays
 in the form of polar jets. This model is the basis for the currently
 favored ``hypernova'' scenario, first proposed by \citet{pacz}, which
 in addition requires the presence of very strong ($\sim 0.1$ T) magnetic 
fields.

There are remarkably few observational probes of failed SNe. The expected rate 
of core-collapse SNe derived from the LF of massive stars predicts many times
fewer SNe than are historically observed in our Galaxy \citep{vdb}, most likely
indicating poor knowledge of the massive star LF. Therefore, it is not possible
to infer from this whether failed SNe exist. Ultimately, with sufficiently
sensitive neutrino detectors, one could directly detect many extragalactic
SNe, and identify the failed ones among them from the lack of a (or very
subluminous) optical counterpart.  Microlensing observations with the {\it
Space Interferometry Mission (SIM)} can directly measure the masses of
isolated BHs and NSs in Galactic bulge fields \citep{GS99,RMF}.  One could
then compare the ratio of NS/BH detections to what would be expected based on,
say, a \citet{scalo} mass function.  The interpretation would be somewhat
complicated by the fact that many NSs receive a large kick at birth, which
could remove a large fraction of them from the bulge.  Nevertheless, given a
sufficiently large sample, and with the information on the kick-velocity
distribution gleaned from the NS transverse velocities (which come out of the
same micolensing observations), it should be possible to reconstruct the
remnant ratio.  Finally, one could examine the ratio of NS/BH X-ray binaries,
where these remnant objects are routinely found.  However, since these are
interacting systems with complex evolutionary histories, it is difficult to
make inferences about the ratio of total populations, and therefore the
production mechanisms, from these very special objects.

Here we propose another probe of failed SNe: BHs in astrometric
binaries.  In astrometric binaries the components are too close to be
resolved, and only the motion of the photocenter is observed. If one component
is invisible (BH or NS), only the motion of the visible component around the
barycenter will be detected. This experiment would be sensitive to detecting
failed SNe that did not undergo some process that would disrupt the
system. \citet{macfadyen} predict that the formation of a collapsar might
always be accompanied by a hypernova explosion, however whether this
necesserrally precludes a BH from remaining in the binary system is uncertain.

In \S~\ref{sec:method}, we discuss the sensitivity of astrometric surveys to
dark companions, focusing specifically on {\it Hipparcos}.  Note that only 
space-based surveys probe enough stars to allow an effective search for 
rare objects. We show that {\it Hipparcos} is sensitive to BH companions of the
great majority of stars in its catalog.  No BH binaries are found, which
potentially places strong limits on the number of failed SNe.  However, since
the initial companion mass function (ICMF) of {\it Hipparcos stars} is
unknown, there is a serious loophole in this argument: one does not know 
whether the absence of BH astrometric binaries reflects the absence of failed 
SNe or the absence of progenitors in binaries.  
In \S~\ref{sec:fame}, we show that
astrometric binary searches using the {\it Full-Sky Astrometric Explorer
(FAME)} can close this loophole.

\section{Astrometric Detection of Dark Companions
\label{sec:method}}

        Consider a binary whose components have masses and luminosities
in the band of astrometric observations $(M,L)$ and $(m,l)$,
respectively.  From Kepler's third law, the semi-major axis, $a$,
is related to the period, $P$, by $[(m+M)/M_\odot](P/\yr)^2=(a/\au)^3$.
If the motion of the photocenter is fit to a Keplerian orbit, the angular
semimajor axis of the photocenter orbit, $\alpha$ (measured in the orbital 
plane), will then be related to the other parameters by
\begin{equation}
{m^3\over M_\odot(m+M)^2}
\biggl({
L\over L + l
}\biggr)^3
=
\biggl({
P\over \yr
}\biggr)^{-2}
\biggl({
D\alpha\over\au
}\biggr)^3,
\label{eqn:kepler1}
\end{equation}
where $D$ is the distance to the system.  The quantities on the rhs of this 
equation can all be measured astrometically.  We will assume that $M$, 
the mass of the more luminous component, can be estimated photometrically
or spectroscopically.  And we will focus on the case in which
the companion is known to be dark (or at least extremely dim compared
to the primary), $l\ll L$.  Under these assumptions, it is straightforward
to determine $m$, the mass of the dark companion, from the astrometric observations.

\begin{figure}
\plotone{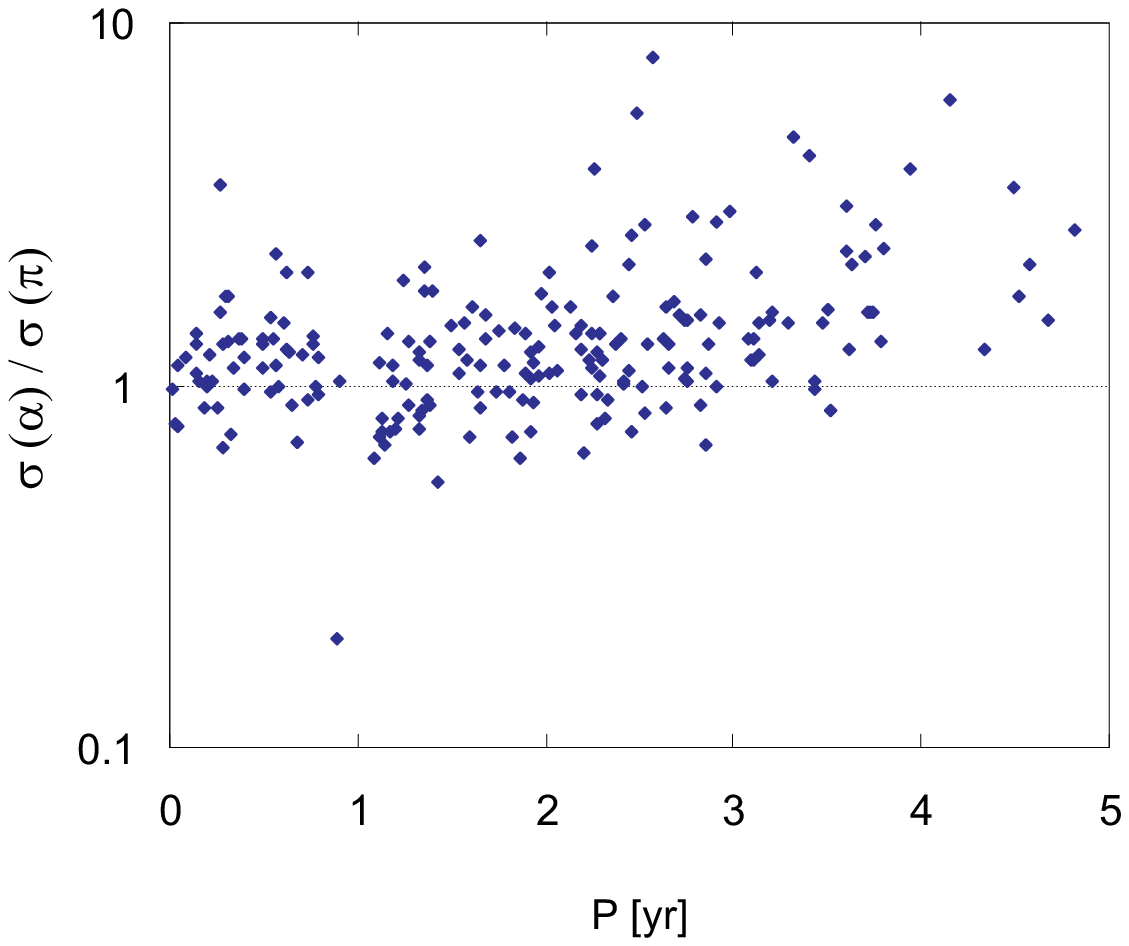}
\caption{\label{fig:one}
The ratio of the error in the photocentric semimajor axis $\sigma_\alpha$
to the parallax error $\sigma_\pi$ as a function of period $P$ for
 astrometric binaries with orbital solutions from the {\it Hipparcos} catalog. 
Plot shows 210 systems (4 lie outside of $y$-axis range.)  As expected
from general arguments, the ratio is typically unity. Note that 
{\it Hipparcos} mission lasted for 3.3 years.
}\end{figure}

In general, $\alpha$ can be measured with approximately the same precision as
the parallax, $\pi$.  Of course this does not hold exactly.  Even for circular
binary orbits, the inclination of the orbit will not match exactly the
ecliptic latitude (i.e., the inclination of the parallactic circle), so there
will be either more or less information about the binary orbit than about the
reflex motion of the Earth's orbit (parallax).  Moreover, for certain binary
orbits, notably edge-on highly eccentric orbits that ``point'' in our
direction, the errors in $\alpha$ will be much larger than the parallax errors
because the binary will show almost no astrometric motion.  Nevertheless, from
the standpoint of making an estimate of the errors for a random ensemble of
binaries, setting $\sigma_\alpha \sim \sigma_\pi$ is a good approximation.
This is confirmed by Figure \ref{fig:one}, where we plot
$\sigma_\alpha/\sigma_\pi$ for astrometric binaries with orbital solutions
(i.e., binaries of type `O') in the Hipparcos catalog \citep[Vol.\ 10]{hip}.
While these fits made use of some auxilliary ground-based spectroscopic
information (mainly to establish the period), or constrained orbits to be
circular, this should not have a major impact on the errors in $\alpha$ for
periods $P\la 3.3\,\yr$, the duration of the mission.  While the figure shows
some scatter, the two errors are roughly equal on average.

\begin{figure}
\plotone{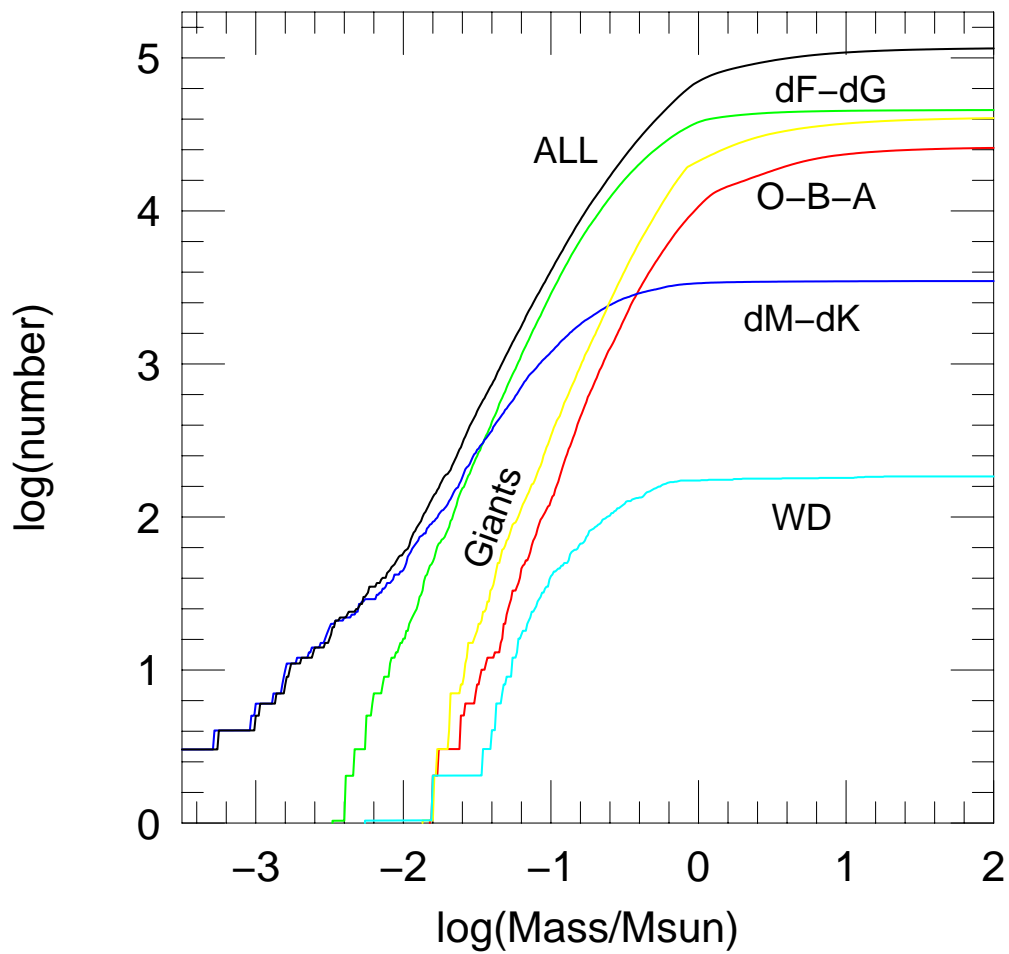}
\caption{\label{fig:two}
Sensitivity of {\it Hipparcos} stars to dark binary companions in 
$P=3.3\,\yr$ orbits as a function of companion mass, according to
stellar type.  Types are broken down (in order of frequency in the
catalog) into F-G dwarfs, giant stars, O-B-A stars, K-M dwarfs, and 
white dwarfs (WDs).  A $5\,\sigma$ signal is required for detection.
Note that WD, NS, and BH companions of mass 0.6, 1.4 and 7 $M_\odot$,
are respectively detectable among 39\%, 68\%, and 89\% of all {\it Hipparcos}
stars.
}\end{figure}

Figure \ref{fig:two} shows the sensitivity ($5\,\sigma$ detection) of
Hipparcos to dark companions as a function of stellar type, i.e., the
number of {\it Hipparcos} stars that can be probed for companions
of a given mass.  These types were assigned based on position in the
color-magnitude diagram when the parallaxes were sufficiently accurate, and
on position in the reduced proper-motion diagram otherwise.  In the latter
case, distances were assigned based on stellar type and color and magnitude.
The figure shows that white dwarf (WD), NS, and BH companions of mass 0.6, 1.4
and 7 $M_\odot$, are respectively detectable among 39\%, 68\%, and 89\% of all
{\it Hipparcos} stars ($N_{Hip}=118,000$).  
For periods of $P=1.5\,\yr$, these
fractions fall to 21\%, 47\%, 52\%.  At $P\sim 1\,\yr$, sensitivity is
seriously compromised by parallax aliasing and at shorter periods the
sensitity falls off rapidly.  On the other hand, for $P\ga 3.3\,\yr$, orbital
solutions become rapidly unstable.  Hence, the sensitivities peak fairly
sharply at $P\sim 3.3\,\yr$.

The overwhelming majority of these {\it Hipparcos} stars are F and G dwarfs,
or giant stars whose progenitors are overwhelmingly F and G dwarfs.  The
frequency of companions per log period for $P\sim 3.3\,\yr$ among such stars
is $df_b/d\log P\sim 7\%$ (\citealt{DM91}).  From the previous paragraph, {\it
Hipparcos} is sensitive to companions over about half a dex in period, $\Delta
\log P\sim 0.5$.  The total number of {\it Hipparcos} stars that were born
with NS/BH progenitor companions in this period range is then,
\begin{equation}
N_{\rm progen} = N_{Hip}{d f_b\over d \log P}\Delta \log P f_{\rm progen}
= 40\,{f_{\rm progen}\over 1\%},
\label{eqn:fprogen}
\end{equation}
where we have normalized the fraction of companions that are NS/BH progenitors
to $f_{\rm progen}=1\%$, in accord with an estimate by \citep{RMF} for their
relative frequency among all stars (both binary and single).  
If a fraction $f_{\rm
fail}$ of these progenitors ended their lives as failed SNe, then there should
be $\sim 40 f_{\rm fail}$ BHs in orbits within the period range covered by
{\it Hipparcos} of which {\it Hipparcos} should be sensitive to $\sim 80\%$ of
them (from the previous paragraph), giving a total of $\sim 30 f_{\rm fail}$ 
BHs. In fact, none of the 235 {\it Hipparcos} astrometric binaries with orbital
solution (188 have $P<3.3\,\yr$) contain a clear BH candidate component.  In 
all cases we find the mass of the companion (if we assume it to be invisible)
either well below the BH range, or of order or smaller than that
of the luminous star,
which implies that the companion is a main sequence star that is fainter than
the primary.

        One would like to use this result to argue that less than 10\% of
massive stars end as failed SNe.  That is, if more than 10\% failed, we would
expect more than 3 BH companions.  Since we find none, the hypothesis would be
ruled out at the 95\% confidence level.

        There are, however, two objections to this line of reasoning.  First,
we do not actually know that at formation the fraction of companions that are
NS/BH progenitors is $f_{\rm progen}=1\%$.  Indeed, there are no observational
constraints on this parameter, and no theoretical reason to believe (or not to
believe) that the fraction of NS/BH progenitors is the same for
G star companions as it is for stars in
the field.  We address this problem in \S~\ref{sec:fame}.

Second, even if this fraction is the same at formation, it could be that the
very process of the failed SNe disrupts the binary.  Certainly, binaries of
this sort will very often be disrupted by an ordinary SNe.  For example,
consider a binary composed of an $M=1\,M_\odot$ and an $m'=8\,M_\odot$ star,
the latter of which ``instaneously'' ejects 83\% of its mass to become an
$m=1.4\,M_\odot$ NS.  Even if the NS receives no kick, the system will become
unbound unless it is near apocenter in a fairly eccentric orbit,
$e>1-2(m+M)/(m'+M)=0.47$. This
eccentricity constraint becomes more severe for larger progenitor masses.
Also, many NSs are known to receive a significant kick, often several hundred
$\kms$, which would certainly disrupt the binary.  However, in a failed SN, a
large fraction of the progenitor would fall back on the BH, so the ratio
$(m+M)/(m'+M)$ would be much larger.  Hence, at least for the relatively less
massive progenitors, the binary would not be disrupted.  It remains possible
that the BH remnant would also receive a strong kick in a failed SN, but since
the mechanism behind the kick is not well understood, this must remain a
matter of speculation.

\begin{figure}
\plotone{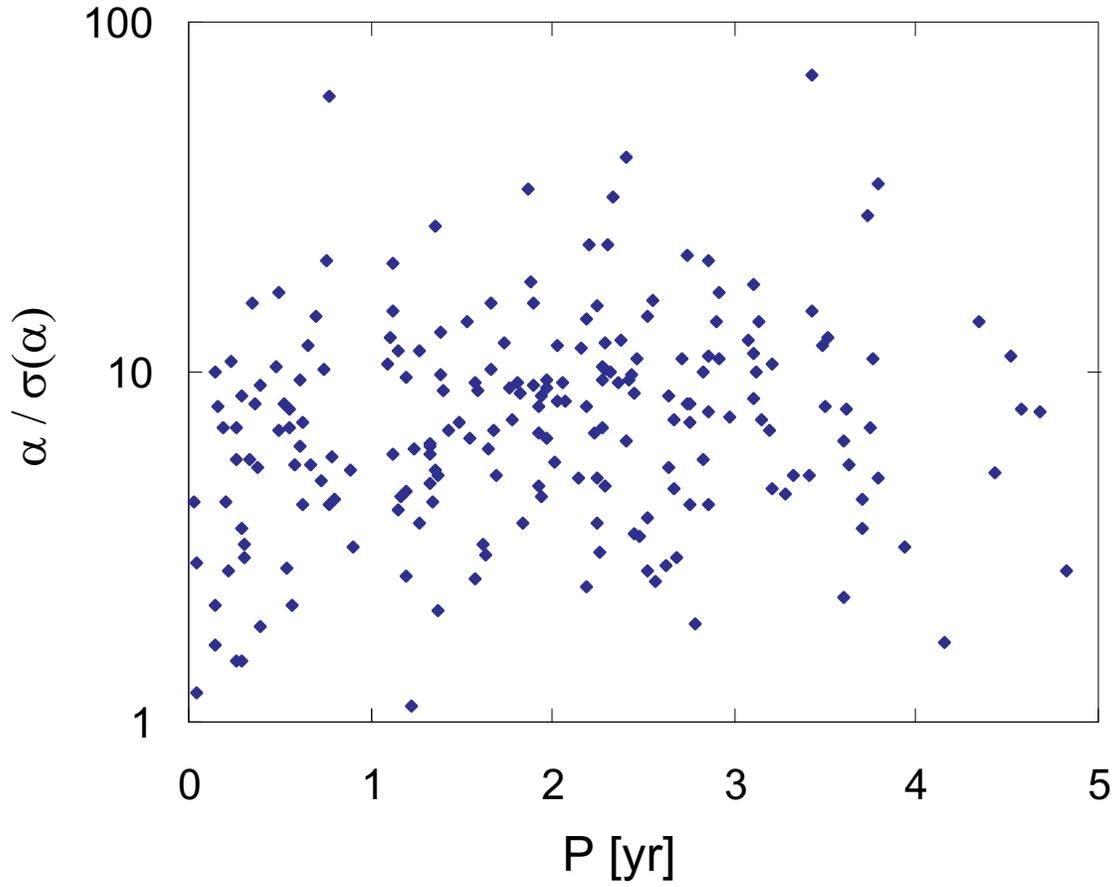}
\caption{\label{fig:three}
Signal-to-noise ratio $\alpha/\sigma_\alpha$ as a function of period $P$
for binaries detected in the {\it Hipparcos} catalog.  From the
form of the distribution, we estimate that the catalog is complete down
to roughly the $5\,\sigma$ detection level.
}\end{figure}

	Finally we note that for the specific case of the {\it Hipparcos}
sample, there is some question as to its real sensitivity.
\citet[Vol.\ 3]{hip} does not quote a specific threshold of detection, i.e.,
the required goodness of the fit, but from Figure \ref{fig:three}, which shows
the distribution of $\alpha/\sigma_\alpha$ as a function of period, we judge
this threshold to be $\sim 5\,\sigma$, i.e.,
the same value we used in estimating the total number of BH companions that
should have been detected.  However, \citet{quist} simulated the number of
detections of $V<7$ MS binaries that {\it Hipparcos} should have made (based on
a Galactic model and the
\citealt{DM91} binary distribution model), and find that
the {\it Hipparcos} catalog should contain between 35\% and 200\% more
binaries with $P<3.3\,\yr$ orbital solutions (type `O') than it actually
does. They
suggest that many binaries for which \citet{hip} finds no orbital solution,
and thus classifies as type `X', or `G', could have produced an orbit if, for
example, the period were known from spectroscopic observations. Careful
reanalysis of {\it Hipparcos} transit data might lead to new orbital
solutions.

\section{The Binary Mass Function
\label{sec:fame}}

While there is some uncertainty as to the fate of binaries containing failed 
SNe, the main problem with deriving robust conclusions from equation
(\ref{eqn:fprogen}) is that the observational constraints on $f_{\rm progen}$,
the fraction of binaries born with a massive companion, a NS/BH progenitor, are
weak.  The main difficulty here is that the ``primaries'' of these systems
(mostly F and G dwarfs and their giant-star descendants -- see Fig.\
\ref{fig:two}) are all about $1\,M_\odot$, whereas the ``secondaries'' of
these systems are substantially more massive.  They are therefore both more
luminous (making the F-G star difficult to detect) and shorter-lived than the
``primaries'' (meaning that they are long gone in a field sample of F-G
stars).  Here we present two complementary astrometric methods to overcome
this difficulty, and show that these can be implemented using {\it FAME}.

We wish to determine the ICMF of G stars as a function of companion mass, $m$,
with some period $P$.  For definiteness, let us consider a period range
$3\,\yr < P <5\,\yr$, and restrict ourselves to one point of the ICMF: B2
stars corresponding to a mass range $11 < m/M_\odot < 15$ and magnitude range
$-2.8<M_V<-2.0$.  What we seek is ratio of formation rate of B2-G binaries (in
this period range) to the formation rate of all G stars.  From this formulation of
the problem, it would seem that one should just survey OB associations and
count the number of B2-G binaries and the total number of G stars.  However,
since the IMF of OB associations may be significantly different from the
disk-averaged IMF, this procedure would produce a biased result.  One must
somehow compare formation rates in the disk as a whole.

\begin{figure}
\plotone{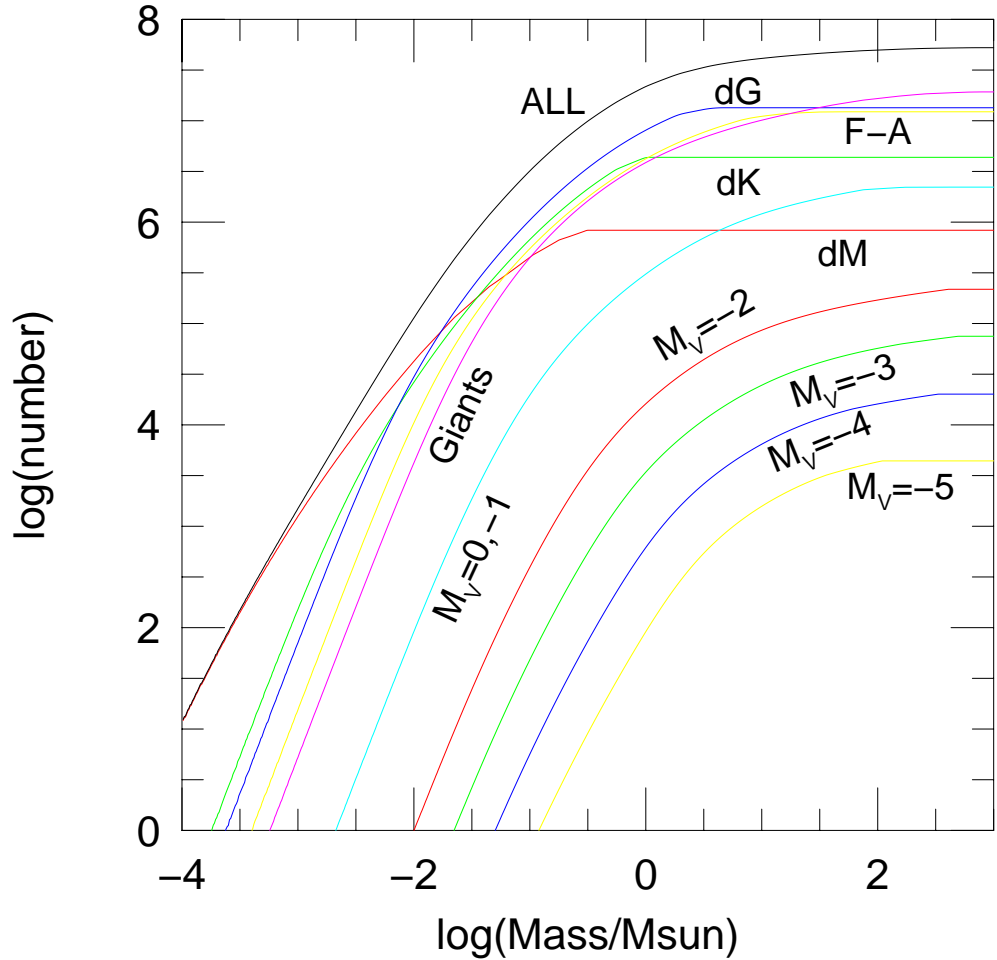}
\caption{\label{fig:bd5}
{\it FAME} sensitivity to binaries with dark (or very dim) companions
in 5-yr periods
as a function of the mass of the lumionous star, for various spectral
types.  The calculation follows that of \citet{sgo}, except that for
early-type stars, the disk scale heights are rescaled in accordance
with \citet{ms}.
}\end{figure}

	Assume for the moment that the star formation rate has been uniform
over the lifetime of the disk.  The ratio we seek, $F^P_{\rm B2|G}$, is then
given by
\begin{equation}
F^P_{\rm B2|G} =
{\Sigma_{\rm B2}\over \Sigma_{\rm G}}\,
{\tau_{\rm G}\over\tau_{\rm B2}}\,
f^P_{\rm G|B2}.
\label{eqn:fp}
\end{equation}
Here $\Sigma_{\rm B2}$ is the column density (number per square parsec)
of B2 stars (averaged
over spiral-arm and inter-arm regions), $\Sigma_{\rm G}$ is the column density
of G stars, $\tau_{\rm G}$ and $\tau_{\rm B2}$ are the lifetimes of G and B2
stars respectively (capped by the age of the disk in the case of late G
stars), and $f^P_{\rm G|B2}$ is the fraction of B2 stars with G companions in
the appropriate period range $P$.  The first two ratios in equation
(\ref{eqn:fp}) are reasonably well known.  The last factor $f^P_{\rm G|B2}$ is
very poorly known but can be measured using {\it FAME}, which will probe a
volume $\sim10,000$ larger than {\it Hipparcos} did. Then using equation 
(\ref{eqn:fp}), one can calculate from the current fraction of B stars with a 
G dwarf companion, the fraction of G dwarfs that were {\it born} with B
companions. Figure \ref{fig:bd5} shows the 
number of stars to which {\it FAME} is sensitive to dark (or dim, 
since a G dwarf is much fainter than a B star) companions with a given mass in 
5 year orbits, for various spectral types.  There are $2\times 10^6$ stars 
with $1.2\,M_\odot \la M \la 8\,M_\odot$ (WD progenitors) for which {\it FAME} 
will be sensitive to companions of $1\,M_\odot$ (G dwarfs), as well as 
$2\times 10^4$ heavier stars $(8\,M_\odot \la M \la 20\,M_\odot)$ (NS or BH 
progenitors).  If the companion rate in this period
range is 7\%/dex, and 5\% of companions are G stars, then {\it FAME} will
detect $\sim 2000$ G star companions of WD progenitors within an octave of
period and $\sim 20$ G star companions of heavier stars.  Thus, the
WD-progenitor ICMF will be mapped out in great detail, and can then be
extended with reasonably good confidence into the higher-mass range of the
progenitors of SNe, luminous or failed.  There will also be a direct
measurement the ICMF in this high mass regime, albeit somewhat crude.

	Of course, the star formation rate has not been uniform over the
lifetime of the disk, but it is straight forward to take account of this
variation by modifying equation (\ref{eqn:fp}).

	This method of determining the ICMF assumes implicitly that $f^P_{\rm
G|B2}$ has not changed from today's value over the age of the disk.  While 
this assumption is
plausible, it can also be partially checked using a different application of
{\it FAME} astrometry, namely by detecting the
remnant object companions themselves.

	In contrast to NS/BH progenitors, the evolution of WD-progenitor
binaries is deterministic provided that the pair is not close enough to
interact during the asymptotic giant branch (AGB) phase.  This is because the
mass loss of the progenitor proceeds on timescales that are long compared to
the period, so that the evolution is adiabatic.  One finds that the final
semimajor axis is $a = a'(M + m')/(M + m)$, where $a'$ is the initial
semimajor axis, and $m'$ is the initial mass.  Hence the initial period $P'$
is related to the final period $P$ by
\begin{equation}
P' = P{(M + m)^2\over (M+m')^2}.
\label{eqn:prelation}
\end{equation}
This means that, while it is not possible to directly probe the time-averaged
ICMF in the NS/BH progenitor regime (because the binaries could have been
disrupted by the SNe), it might be 
possible to probe it in the WD-progenitor regime.

	A significant problem in determining the ICMF for G stars is that when
a $0.6\,M_\odot$ companion to a G dwarf is discovered, one does not
immediately know whether this companion is a WD or an M dwarf (with $M_V\sim
8$).  Neither has much luminosity compared to the G dwarf and so in neither
case would the mass determination be significantly affected (see eq.\
[\ref{eqn:kepler1}]).  By the same token, however, there would be no obvious
signatures that the companion was one type or the other. 
It is possible that with precision photometry one could detect the IR
excess due to the M dwarf.  High signal-to-noise ratio spectroscopy
could certainly detect the M dwarf if it were there.  The scope of the
spectroscopy project would be significantly reduced if one surveyed K dwarfs
rather than G dwarfs because the K/M magnitude difference is substantially
smaller than the G/M difference.  Figure \ref{fig:bd5} shows that {\it FAME}
will be senstive to WD companions of $10^6$ K dwarfs and $2\times 10^6$ G
dwarfs, which is certainly enough to obtain a large sample of WD/dG-dK
binaries.

	The major limitation of this method comes from the fact that during 
its 5-yr mission, {\it FAME} 
can obtain accurate mass measurements only for $P< 5\,\yr$.  According
to equation (\ref{eqn:prelation}), the periods $P'$ of the progenitor systems
were substantially shorter than the periods $P$ 
of their present-day descendants.
Specifically, for G dwarf primaries, we have $P'(M+m')^2 < 12.8 M_\odot^2\,\rm
yr$.  If the periods were too short, then during the AGB phase, the binary 
would have suffered mass transfer and its evolution would have taken 
a complex course. If we assume that 
no mass transfer occurs for $a'>1.5\,\au$ (the 
exact value is model dependent), 
then this condition implies $(M+m')(a'/\au) < 12.8^{2/3}M_\odot$, or
$m'<2.6\,M_\odot$.  Hence only the lower-mass 
WD-progenitor population is probed.
Moreover, in this mass range, the WD mass is only a very weak function of the
progenitor mass (hence the peakiness of the WD mass function).  Given both the
astrometric errors and the errors in the photometric masses of the G star
primaries, it seems unlikely that one could obtain much more detail than the
total number of WDs as a function of period.  Hence one would really obtain
only a single point beyond the usual G dwarf ICMF (i.e., of secondaries that 
are
fainter and lower-mass than the primary).  Nevertheless, equal-mass is the only
natural scale in this problem.  Thus, if this direct determination of the
time-averaged ICMF tracked the ICMF measured from present day companions of
early-type stars across the equal-mass boundary, it would lend credence to the
latter ICMF measurement at higher masses.

	If GAIA ultimately flies, then one could extend the orbits 
initially mapped out by FAME to effectively cover periods of $\sim 15$
years, which would allow one to probe WDs that have much larger progenitor
masses.

\section{Conclusion}

	The notion that some massive stars undergo a collapse without 
producing a SN first appeared as a failure to produce SN explosions in 
hydrodynamical simulations, but later gained ground as a possible mechanism 
behind GRBs. These theoretical considerations were left without any empirical 
evidence. If in binary systems these failed SNe collapse into BHs
without disrupting the binaries, one might be able detect such systems 
astrometrically. To this end we suggest using {\it FAME} since it will observe 
vast number of stars with great precision. To estimate the rate of SNe that 
fail, it is first neccessary to estimate the number of binaries that were born 
consisting of a G dwarf (a typical star {\it FAME} will observe), and a 
massive, short-lived B star (NS/BH progenitor). We can derive this number 
using {\it FAME} detections of either currently existing G+B pairs, or 
by extrapolating from the number of G+WD binaries. {\it FAME} will be able
to probe $\sim 4\times 10^7$ stars for BH companions.  The result of this 
experiment would either discover a new phenomenon (BH collapsar), or place 
stringent limits on SN and GRB models.

\acknowledgements
This work was supported in part by 
JPL contract 1226901.

\appendix


\end{document}